\documentclass[11pt, a4paper]{article}

\usepackage[UKenglish]{babel}
\usepackage{amsmath, amsfonts, graphicx, float, hyperref}

\allowdisplaybreaks

\hypersetup{
    colorlinks=true,
    linkcolor=black,
    citecolor=black,
    filecolor=black,
    urlcolor=black
}

\DeclareMathOperator{\re}{Re}

\DeclareMathOperator{\const}{const}

\begin{document}

\pagenumbering{roman}

\begin{titlepage}

\begin{flushright}
{\bf \today} \\
DAMTP-2012-16\\

\end{flushright}

\begin{centering}

\vspace{3cm}

{\Large{\bf On the BPS Spectrum at the Root of the Higgs Branch}}

\vspace{1cm}

Nick Dorey and Kirill Petunin
\\[1cm]
DAMTP, Centre for Mathematical Sciences,\\
University of Cambridge, Wilberforce Road,\\
Cambridge, CB3 0WA, UK
\\[1cm]

\vspace{1cm}

{\bf \large Abstract} \\
\vspace{0.25cm}
We study the BPS spectrum and walls of marginal stability of the $\mathcal{N}=2$ supersymmetric theory in four dimensions with gauge group $SU(n)$ and $n\le N_{f}<2n$ fundamental flavours at the root of the Higgs branch.
The strong-coupling spectrum of this theory was conjectured in hep-th/9902134 to coincide with that of the two-dimensional supersymmetric $\mathbb{CP}^{2n-N_{f}-1}$ sigma model.
Using the Kontsevich--Soibelman wall-crossing formula, we start with the conjectured strong-coupling spectrum and extrapolate it to all other regions of the moduli space.
In the weak-coupling regime, our results precisely agree with the semiclassical analysis of hep-th/9902134: in addition to the usual dyons, quarks, and $W$ bosons, if the complex masses obey a particular inequality, the resulting weak-coupling spectrum includes a tower of bound states consisting of a dyon and one or more quarks.
In the special case of $\mathbb{Z}_{n}$-symmetric masses, there are bound states with one quark for odd $n$ and no bound states for even $n$.
\end{centering}

\end{titlepage}

\pagenumbering{arabic}


\section{Introduction}

\paragraph{}

We consider the BPS spectrum of the $\mathcal{N}=2$ SQCD with gauge group $SU(n)$ and $n\le N_{f}<2n$ fundamental flavours at the root of the Higgs branch in four dimensions.
The central charge in this theory was shown in \cite{Dorey 1998} to be the same as in the $\mathcal{N}=(2,2)$ supersymmetric $\mathbb{CP}^{2n-N_{f}-1}$ sigma model in two dimensions with twisted mass terms (with an appropriate identification of parameters).
On this basis, the BPS spectra of the two theories were also conjectured to coincide~\footnote{
By this, we mean that there is a single multiplet of $\mathcal{N}=(2,2)$ SUSY in the two-dimensional theory associated with each massive multiplet of $\mathcal{N}=2$ SUSY in the four-dimensional theory.
With an appropriate identification of parameters, the masses of corresponding states agree exactly.
For more details see \cite{Dorey 1998, DHT}.
}.
Additional support for the conjecture was provided \cite{DHT}, and an explanation for the coincidence of the spectra in terms of vortex strings was given in \cite{SY, Hanany Tong 2}.
In this paper, we present further evidence for this conjecture using the Kontsevich--Soibelman wall-crossing formula for the four-dimensional theory.
The finite set of BPS states of the two-dimensional theory in the strong-coupling regime is known \cite{Witten phases, CV}.
Assuming that the strong-coupling spectrum of the four-dimensional theory is indeed the same, we find the walls of marginal stability and, employing the wall-crossing formula, extrapolate the spectrum to other regions of the moduli space.
In the weak-coupling region, we recover the complete semiclassical spectrum derived in \cite{DHT}, thus confirming our starting assumption.
The conclusions of a forthcoming analysis of the spectrum of the corresponding two-dimensional theory \cite{BSY2} are fully consistent with our results.

\paragraph{}

From our analysis, the following general picture emerges.
For a given magnetic charge, there is a (``primary'') wall separating the strong-coupling region from the rest of the moduli space.
Outside this wall, the spectrum expands and includes an infinite (``primary'') tower of dyons as well as quarks and $W$ bosons.
In addition, we show that if a particular condition on the complex masses is satisfied, there is one extra (``secondary'') tower of bound states consisting of a dyon and one or more quarks.
Unlike the primary case, the wall-crossing formula shows that all the states in the extra tower cannot be created at a single wall.
Rather, for every bound state in the tower, there is a corresponding (``secondary'') wall.
We also show that each secondary wall separates the primary wall from the weak-coupling region, so that all walls must be traversed in passing between strong and weak coupling.

\paragraph{}

A particular configuration of $\mathbb{Z}_{n}$-symmetric masses, when all $n$ masses form a regular polygon in the complex plane, can be analysed more explicitly: we find that there exists one secondary tower of bound states with one quark for odd $n$ and no bound states for even $n$.

\paragraph{}

Let us introduce our conventions in the four-dimensional theory: $N_{f}=n+\tilde n$ is the total number of flavours, $\vec q_{e}$ and $\vec q_{m}$ are the vectors of electric and magnetic charges with $n$ components (counted by $I$), $\vec S$ is the vector of flavour charges with $n+\tilde n$ components (counted by $i$).
The central charge is given by
\begin{equation}
\label{central charge general}
Z_{(\vec q_{e},\vec q_{m},\vec M)} = \vec a\vec q_{e}+\vec a_{D}\vec q_{m}+\vec S\vec M = \sum_{I=0}^{n-1} \left( a^{I}q_{e\,I}+a_{D\,I}q_{m}^{I} \right)+\sum_{i=0}^{N_{f}-1}S_{i}M_{i}
\end{equation}
where $\vec a$ is the vacuum expectation value, $\vec a_{D}$ is its magnetic dual, $\vec M$ is the vector of flavour masses.
We divide $\vec S$ and $\vec M$ into two pieces: $\vec s$ and $\vec m$ contain the first $n$ components corresponding to the massless quarks at the root of the Higgs branch, $\vec{\tilde s}$ and $\vec{\tilde m}$ contain the remaining $\tilde n$ components; we distinguish the remaining $\tilde n$ flavour components by putting a tilde above their masses, charges, and indices.
The root of the Higgs branch is determined by setting $\vec a=-\vec m$; analogously, we define a magnetic dual mass $\vec m_{D}=-\vec a_{D}(\vec{a}=-\vec{m})$.
Therefore, the central charge (\ref{central charge general}) reduces to
\begin{equation}
\label{central charge root}
\begin{aligned}
& Z_{(\vec\gamma_{e},\vec\gamma_{m},\vec{\tilde s})} = \vec m\vec\gamma_{e}+\vec m_{D}\vec\gamma_{m}+\vec{\tilde s} \, \vec{\tilde m} = \sum_{I=0}^{n-1} \left( m^{I}\gamma_{e\,I}+m_{D\,I}\gamma_{m}^{I} \right)+\sum_{\tilde i=0}^{\tilde n-1}s_{\tilde i}m_{\tilde i}
\,,
\\
& \vec\gamma_{e} = -\vec q_{e}+\vec s
\,,
\quad
\vec\gamma_{m} = -\vec q_{m}
\,.
\end{aligned}
\end{equation}
Now, for each BPS state, the complete set of (electric, magnetic, and flavour) charges is $\gamma=(\vec\gamma_{e},\vec\gamma_{m},\vec{\tilde s}\,)$; if $\vec{\tilde s}=\vec 0$, we will omit the third entry.

\paragraph{}

Our approach is mainly based on the Kontsevich--Soibelman wall-crossing formula \cite{KS}.
For a given charge $\gamma=(\vec\gamma_{e},\vec\gamma_{m})$, we define the Kontsevich--Soibelman operator acting on the so-called Darboux coordinates $\mathcal{X}_{\beta}$ (for any charge $\beta$) as
\begin{equation}
\label{KS operator}
\mathcal{K}_{\gamma} \quad \colon \quad
\mathcal{X}_{\beta} \to \mathcal{X}_{\beta} \left( 1-\sigma(\gamma)\mathcal{X}_{\gamma} \right)^{\langle\beta,\gamma\rangle}
\end{equation}
where for any pair of charges, $\alpha=(\vec\alpha_{e},\vec\alpha_{m})$ and $\beta=(\vec\beta_{e},\vec\beta_{m})$, the symplectic product is defined as
\begin{equation}
\label{symplectic product}
\langle(\vec\alpha_{e},\vec\alpha_{m}),(\vec\beta_{e},\vec\beta_{m})\rangle =
-\vec\alpha_{e}\vec\beta_{m}+\vec\alpha_{m}\vec\beta_{e}
\,,
\end{equation}
and the quadratic refinement is defined as
\begin{equation}
\label{quadratic refinement}
\sigma(\gamma) = (-1)^{\vec\gamma_{e}\vec\gamma_{m}}
\,.
\end{equation}
Let $\Gamma(\vec M)$ be the set of BPS states, depending on the set of masses $\vec M$.
We associate the following operator to each point $\vec M$ in the moduli space:
\begin{equation}
S = \prod_{\gamma\in\Gamma(\vec M)} \mathcal{K}_{\gamma}^{\Omega(\gamma,\vec M)} = \const
\end{equation}
where $\Omega(\gamma,\vec M)$ is the degeneracy of the BPS state with charge $\gamma$; all operators (i.e., their BPS rays) are ordered clockwise (equivalently, their central charges as complex vectors are ordered counterclockwise).
The statement of the wall-crossing formula is that, although the spectrum and the ordering of operators change across the moduli space, the resulting product $S$ is constant.
In principle, knowing the set of charges on one side of the wall of marginal stability, we can compute them on the other side of the wall \cite{GMN}.
In practice, this proceeds via the use of known identities, such as (\ref{WCF pure}) and (\ref{pentagon root}) below, which apply for specific values of the symplectic product of the two states whose central charges become aligned at the wall.


\section{The wall at strong coupling}

\paragraph{}

Our starting point is the strong-coupling spectrum of the theory.
As mentioned above, we start by assuming the spectrum implied by the 2d/4d conjecture of \cite{Dorey 1998, DHT}.
For the moment, we consider only the BPS states corresponding to kinks interpolating between two neighbouring vacua in the two-dimensional theory.
Without loss of generality, we can set the magnetic charge to be equal to $(-1,1,0,0,\dots)$.
There are exactly $n$ such states (plus charge conjugates).
Electric charges are determined only up to a fixed shift \cite{DHT}; for our purposes, it is convenient to choose this shift so that the charges of the states are
\begin{equation}
\label{spectrum strong root}
\begin{aligned}
\pm\gamma_{1} & = \pm ((1,0,0,0,0,\dots),(-1,1,0,0,0,\dots))
\,,
\\
\pm\gamma_{2} & = \pm ((0,1,0,0,0,\dots),(-1,1,0,0,0,\dots))
\,,
\\
\pm\gamma_{3} & = \pm ((0,0,1,0,0,\dots),(-1,1,0,0,0,\dots))
\,,
\\
& \dots
\end{aligned}
\end{equation}
The moduli space includes Argyres--Douglas points \cite{Argyres Douglas} located on the boundary of the strong-coupling region.
One of these states becomes massless at each of these points.

\paragraph{}

\begin{figure}[ht]
\centering
\includegraphics[width=65mm]{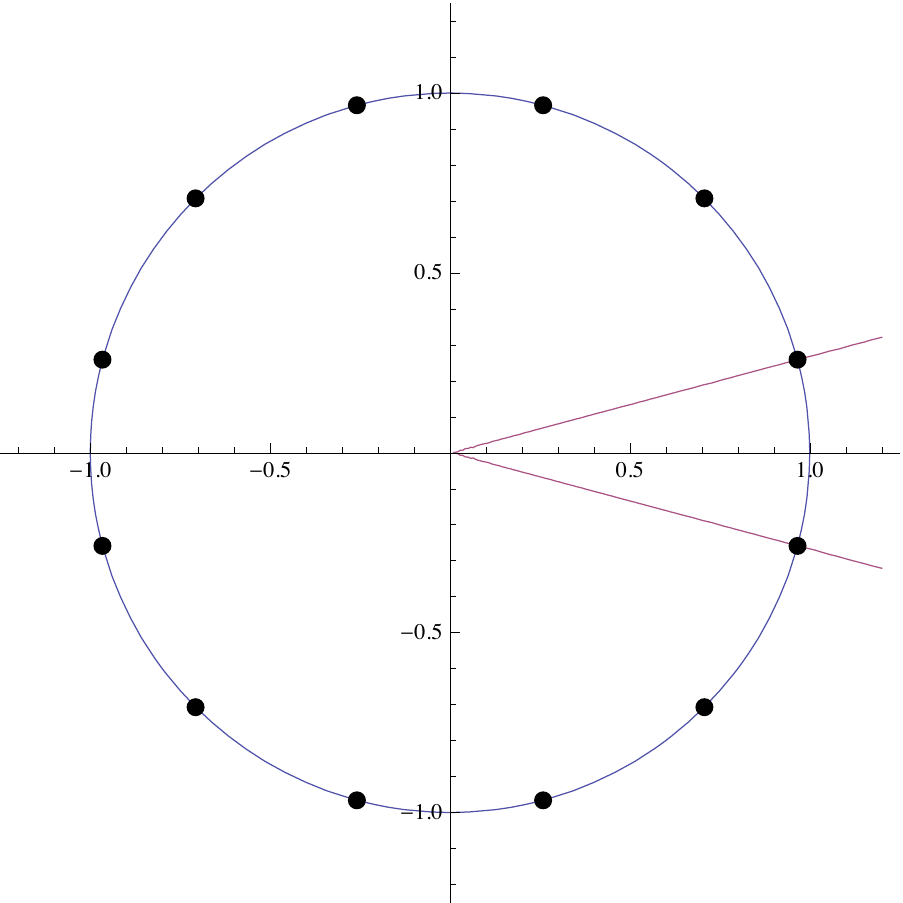}
\caption{
Argyres--Douglas points for $\mathbb{Z}_{12}$-symmetric masses in the $m_{0}$ plane.
}
\label{fig: massless states root}
\end{figure}

Consider the case of $\mathbb{Z}_{n}$-symmetric masses.
We set
\begin{equation}
m_{I} = m_{0} \, \exp\frac{2\pi iI}{n}
\,,
\end{equation}
where $m_{0}$ remains as a free parameter which interpolates between strong and weak coupling.
Then, all central charges and walls of marginal stability can be determined as functions of $m_{0}$, and the masses automatically obey
\begin{equation}
\sum_{I=0}^{n-1} m_{I} = 0
\,.
\end{equation}
A convenient expression for the magnetic dual masses in the $\mathbb{Z}_{n}$-symmetric case is \cite{BSY}~\footnote{
Although the analysis of \cite{BSY} is for the corresponding two-dimensional theory, the resulting formula for the central charge is identical to that of the four-dimensional theory at the Higgs branch root \cite{Dorey 1998}.
}
\begin{equation}
m_{D\,I} = e^{2\pi iI/n}
\left(
n\sqrt[n]{m_{0}^{n}+\Lambda^{n}} +
\sum_{k=0}^{n-1} m_{0}e^{2\pi ik/n} \log\frac{\sqrt[n]{m_{0}^{n}+\Lambda^{n}}-m_{0}e^{2\pi ik/n}}{\Lambda}
\right)
\end{equation}
where the branch is fixed by requiring that for $x\in\mathbb{R}_{+}$, $\sqrt[n]{x}\in\mathbb{R}_{+}$ and $\log x\in\mathbb{R}$, as in \cite{BSY}.
Then, the Argyres--Douglas points, where the strong-coupling states (\ref{spectrum strong root}) become massless, are located at \cite{BSY}
\begin{equation}
m_{0} = \Lambda \, \exp\frac{\pi i(2j+1)}{n}
\,,
\quad
j \in \mathbb{Z}
\end{equation}
(figure \ref{fig: massless states root}).
As explained in \cite{Olmez Shifman}, for $\mathbb{Z}_{n}$-symmetric masses, it is sufficient to consider $m_{0}$ belonging to the sector between two neighbouring Argyres--Douglas points, $\Lambda e^{\pi i/n}$ and $\Lambda e^{-\pi i/n}$, where $\gamma_{1}$ and $\gamma_{2}$ from (\ref{spectrum strong root}) are massless.

\paragraph{}

\begin{figure}[ht]
\centering
\includegraphics[width=65mm]{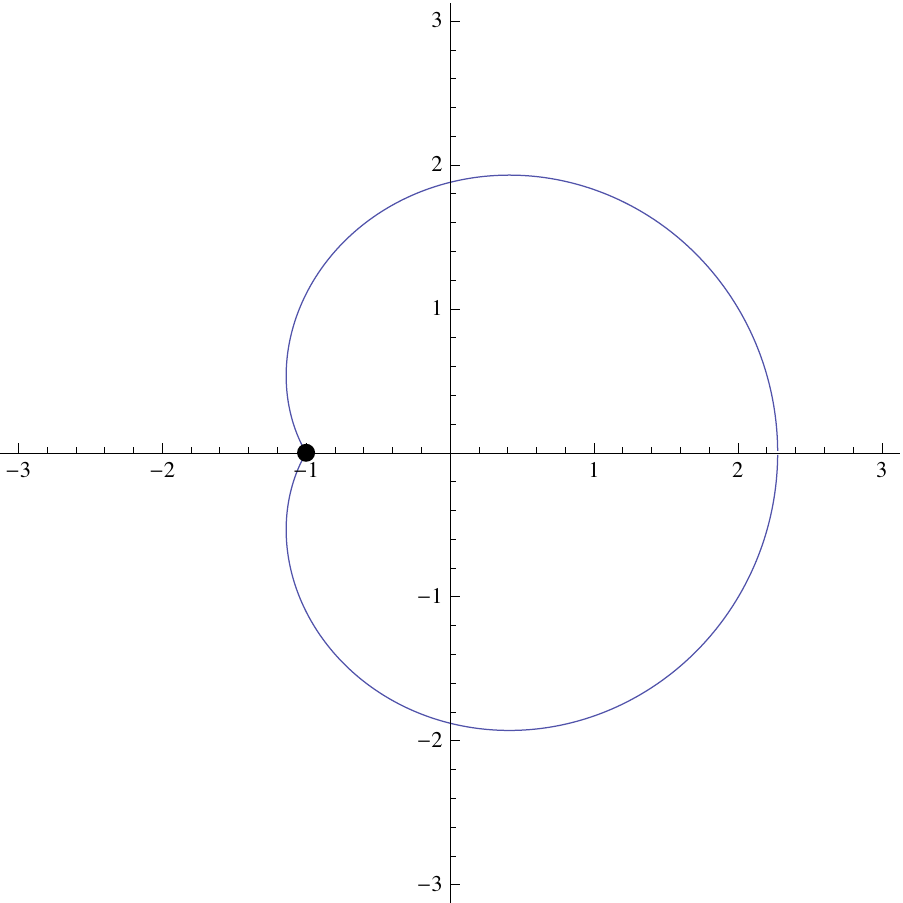}
\qquad
\includegraphics[width=65mm]{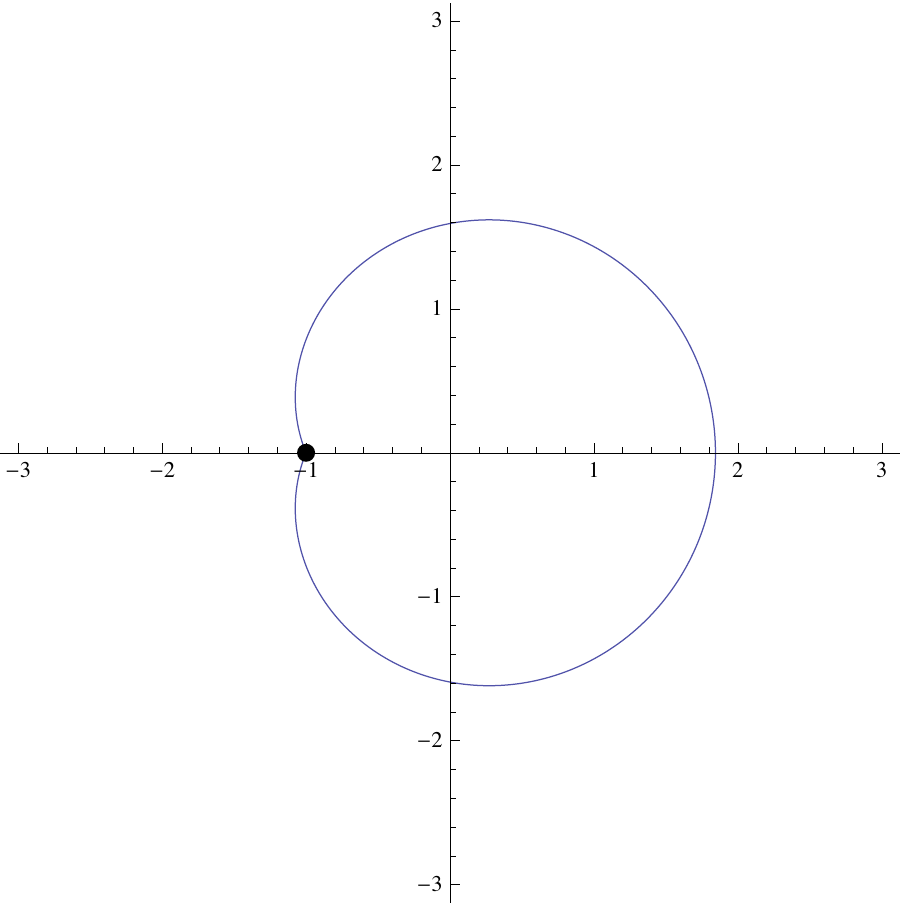}
\caption{
The primary walls of marginal stability for $\mathbb{Z}_{2}$- and $\mathbb{Z}_{3}$-symmetric masses \cite{Shifman Vainshtein Zwicky, BSY}.
}
\label{fig: walls primary root}
\end{figure}

Let us find out how the spectrum changes when $\vec M$ crosses the primary wall of marginal stability, where the central charges of the first two dyons in (\ref{spectrum strong root}), $\gamma_{1}$ and $\gamma_{2}$, become aligned:
\begin{equation}
\frac{Z_{\gamma_{1}}}{Z_{\gamma_{2}}} \in \mathbb{R}_{+}
\end{equation}
(figure \ref{fig: walls primary root}).
Using the wall-crossing formula, we can compute the spectrum on the external side of the wall.
The symplectic product is
\begin{equation}
\langle\gamma_{1},\gamma_{2}\rangle = -2
\,,
\end{equation}
hence, the relevant wall-crossing formula is essentially the same as the formula relating the strong- and weak-coupling spectra of the pure $SU(2)$ theory in four dimensions~\footnote{
This reflects the fact that all the states involved have charges contained in an $SU(2)$ subgroup of the gauge group.
} \cite{KS}:
\begin{equation}
\label{WCF pure}
\mathcal{K}_{-\gamma_{2}} \mathcal{K}_{\gamma_{1}} =
\mathcal{K}_{\gamma_{1}} \mathcal{K}_{2\gamma_{1}-\gamma_{2}}
\mathcal{K}_{3\gamma_{1}-2\gamma_{2}}
\mathcal{K}_{4\gamma_{1}-3\gamma_{2}} \dots
\mathcal{K}_{\gamma_{1}-\gamma_{2}}^{-2} \dots
\mathcal{K}_{3\gamma_{1}-4\gamma_{2}}
\mathcal{K}_{2\gamma_{1}-3\gamma_{2}}
\mathcal{K}_{\gamma_{1}-2\gamma_{2}} \mathcal{K}_{-\gamma_{2}} \,.
\end{equation}
In our notations, the part of the wall-crossing formula that changes across the wall takes the following form:
\begin{equation}
\label{WCF root}
\begin{aligned}
\mathcal{K}_{-((0,1),(-1,1))} \mathcal{K}_{((1,0),(-1,1))}
& =
\mathcal{K}_{((1,0),(-1,1))} \mathcal{K}_{((2,-1),(-1,1))} \mathcal{K}_{((3,-2),(-1,1))} \mathcal{K}_{((4,-3),(-1,1))}
\\
\dots \mathcal{K}_{((-1,1),(0,0))}^{-2} & \dots
\mathcal{K}_{-((-3,4),(-1,1))} \mathcal{K}_{-((-2,3),(-1,1))} \mathcal{K}_{-((-1,2),(-1,1))} \mathcal{K}_{-((0,1),(-1,1))}
\end{aligned}
\end{equation}
where we display only the first two components of electric and magnetic charges, the others being equal to zero.
This relation shows that the spectrum outside the wall consists of a tower of dyons and a finite number of quarks and $W$ bosons with charges
\begin{equation}
\begin{aligned}
& \pm ((-\nu+1,\nu, 0,0,\dots),(-1,1,0,0,\dots))
\,,
\\
& \pm ((-1,1,0,0,\dots),(0,0,0,0,\dots))
\,.
\end{aligned}
\end{equation}


\section{Bound states}

\paragraph{}

In fact, the complete BPS spectrum is not limited to the primary tower of states found above: depending on the values of masses, there can also be secondary towers of bound states formed by a dyon and $p$ quarks \cite{DHT}.
Creation (or, conversely, destruction) of these extra states is described by the pentagon formula:
\begin{equation}
\label{pentagon root}
\mathcal{K}_{\gamma_{1}}\mathcal{K}_{\gamma_{2}} =
\mathcal{K}_{\gamma_{2}}\mathcal{K}_{\gamma_{1}+\gamma_{2}}\mathcal{K}_{\gamma_{1}}
\,,
\quad
\forall \ \langle\gamma_{1},\gamma_{2}\rangle = \pm 1
\end{equation}
where the new state $\gamma_{1}+\gamma_{2}$ is created from $\gamma_{1}$ and $\gamma_{2}$ where one of the initial states is a quark, and the other one is either a dyon or a bound state consisting of a dyon and $p-1$ quarks.
This process occurs when
\begin{equation}
\label{secondary wall root}
\frac{Z_{\gamma_{1}}}{Z_{\gamma_{2}}} \in \mathbb{R}_{+}
\,.
\end{equation}
We will find the resulting secondary walls and prove that they are located outside the primary wall and have to be crossed as the VEV moves from strong to weak coupling.
The restriction on the wedge-product of the two interacting states in (\ref{pentagon root}) allows us to determine which states can combine to form a bound state if the corresponding secondary wall exists.

\paragraph{}

Starting with the states constructed in the previous section, when $\tilde n=0$, we can see that there can be two possible types of creation processes, both leading to the same set of new states:
\begin{equation}
\label{first process root}
\begin{aligned}
1:
\quad
((-\nu+1,\nu,0,0,0,\dots),(-1,1,0,0,0,\dots)) + ((-1,0,1,0,0,\dots),(0,0,0,0,0,\dots))
\\
\leftrightarrow
((-\nu,\nu,1,0,0,\dots),(-1,1,0,0,0,\dots))
\,,
\end{aligned}
\end{equation}
\begin{equation}
\label{second process root}
\begin{aligned}
2:
\quad
((-\nu,\nu+1,0,0,0,\dots),(-1,1,0,0,0,\dots)) + ((0,-1,1,0,0,\dots),(0,0,0,0,0,\dots))
\\
\leftrightarrow
((-\nu,\nu,1,0,0,\dots),(-1,1,0,0,0,\dots))
\,.
\end{aligned}
\end{equation}
These are the bound states formed by a dyon and one quark.
Explicitly, the walls of marginal stability (\ref{secondary wall root}) for these processes are determined by
\begin{align}
\label{first wall root}
1:
\quad
\frac{-m_{0}+m_{2}}{(-\nu+1)m_{0}+\nu m_{1}-m_{D\,0}+m_{D\,1}} & \in \mathbb{R}_{+}
\,,
\\
2:
\quad
\label{second wall root}
\frac{-m_{1}+m_{2}}{-\nu m_{0}+(\nu+1)m_{1}-m_{D\,0}+m_{D\,1}} & \in \mathbb{R}_{+}
\,.
\end{align}

\paragraph{}

For general $N_{f}$, there are additional bound states involving the remaining $\tilde n$ flavours:
\begin{equation}
\begin{aligned}
((-\nu+1,\nu,0,0,0,\dots),(-1,1,0,0,0,\dots)) + ((-1,0,0,0,0,\dots),(0,0,0,0,0,\dots),(1,0,0,\dots))
\\
\leftrightarrow
((-\nu,\nu,0,0,0,\dots),(-1,1,0,0,0,\dots),(1,0,0,\dots))
\,,
\end{aligned}
\end{equation}
\begin{equation}
\begin{aligned}
((-\nu,\nu+1,0,0,0,\dots),(-1,1,0,0,0,\dots)) + ((0,-1,0,0,0,0,\dots),(0,0,0,0,0,\dots),(1,0,0,\dots))
\\
\leftrightarrow
((-\nu,\nu,0,0,0,\dots),(-1,1,0,0,0,\dots),(1,0,0,\dots))
\,.
\end{aligned}
\end{equation}
They are completely analogous to the ones above: the walls of marginal stability for these processes can be obtained by changing $m_{2}$ to $\tilde m_{0}$ in the previous formulae.

\paragraph{}

As has been discussed above, there can also be bound states formed by a dyon and $p$ quarks:
\begin{equation}
\label{bound state general root}
\begin{aligned}
& ((-\nu+1+p,\nu,j_{3},j_{4},j_{5},\dots),(-1,1,0,0,0,\dots),(\tilde j_{\tilde 1},\tilde j_{\tilde 2},\tilde j_{\tilde 3},\dots))
\,,
\\
& j_{i}(j_{i}-1) = \tilde j_{\tilde i}(\tilde j_{\tilde i}-1) = 0
\,,
\quad
p + \sum_{i=2}^{n-1} j_{i} + \sum_{\tilde i=0}^{\tilde n-1} \tilde j_{\tilde i} = 0
\,.
\end{aligned}
\end{equation}
They exist if starting with the strong-coupling spectrum and moving into the weak-coupling region, $|p|$ different secondary walls of marginal stability (\ref{secondary wall root}) are crossed.

\paragraph{}

We need to find out if the processes (\ref{first process root}) and (\ref{second process root}), which we rewrite as
\begin{equation}
\label{first process root 2}
\begin{aligned}
1:
\quad
d_{1} + q_{1}
\leftrightarrow
b
\,,
\end{aligned}
\end{equation}
\begin{equation}
\label{second process root 2}
\begin{aligned}
2:
\quad
d_{2} + q_{2}
\leftrightarrow
b
\,,
\end{aligned}
\end{equation}
actually take place when the masses move from strong to weak coupling: to do this, we should check whether the secondary walls (\ref{secondary wall root}) are crossed, i.e., if the following conditions are satisfied somewhere outside the primary wall of marginal stability:
\begin{equation}
\label{first wall root 2}
1:
\quad
\arg Z_{d_{1}} = \arg Z_{q_{1}}
\,,
\end{equation}
\begin{equation}
\label{second wall root 2}
2:
\quad
\arg Z_{d_{2}} = \arg Z_{q_{2}}
\,.
\end{equation}
Note that $Z_{q_{j}}$ ($j=1$ or $j=2$) is independent of the region in the moduli space, and $\arg Z_{d_{j}}$ changes continuously between the primary wall and the weak-coupling region, therefore, (\ref{first wall root 2}) (with $j=1$) and (\ref{second wall root 2}) (with $j=2$) are satisfied somewhere if in the complex plane, $Z_{q_{j}}$ lies between the values of $Z_{d_{j}}$ at the primary wall and in the weak-coupling limit.

\paragraph{}

To check if this is the case, it is convenient to start at the Argyres--Douglas point where either $\gamma_{1}$ or $\gamma_{2}$ in (\ref{spectrum strong root}) becomes massless.
Consider (\ref{first wall root 2}) first.
For $\nu>0$ and for $\nu\le 0$, we start at the points $s_{1}$ and $s_{2}$ where $\gamma_{1}$ and $\gamma_{2}$ in (\ref{spectrum strong root}) are massless, respectively.
Near these points, the corresponding central charges of dyons can be approximated as
\begin{equation}
\label{central charge dyon initial root}
(Z_{d_{1}})_{s_{1}} \simeq \nu(-m_{0}+m_{1})
\,,
\quad
(Z_{d_{1}})_{s_{2}} \simeq (\nu-1)(-m_{0}+m_{1})
\,.
\end{equation}
Then, we continuously move the masses into the semiclassical region, where
\begin{equation}
\label{central charge dyon final root}
(Z_{d_{1}})_{w} \simeq i(-m_{0}+m_{1})
\,.
\end{equation}
From (\ref{central charge dyon initial root}) and (\ref{central charge dyon final root}), we have
\begin{equation}
\label{central charge dyon change root}
\lim_{g_{\rm eff}\to 0} \arg\frac{(Z_{d_{1}})_{w}}{(Z_{d_{1}})_{s_{l}}} = (-1)^{l-1}\frac{\pi}{2}
\,,
\quad
\left| \arg\frac{(Z_{d_{1}})_{w}}{(Z_{d_{1}})_{s_{l}}} \right| < \frac{\pi}{2}
\,,
\end{equation}
where the inequality is strict for any $g_{\rm eff}$ because the central charge receives corrections from its electric components at weak coupling.
We can consider (\ref{second wall root 2}) analogously: (\ref{first process root}) and (\ref{second process root}) are related by changing $\nu\to-\nu$ and swapping $\gamma_{e\,0}\leftrightarrow\gamma_{e\,1}$.
Comparing $Z_{q_{j}}$ with (\ref{central charge dyon initial root}) and (\ref{central charge dyon final root}) and using (\ref{central charge dyon change root}), we conclude that the walls (\ref{first wall root}, \ref{second wall root}) exist when
\begin{equation}
\label{first bound state constraint root}
\begin{aligned}
1:
\quad
& \nu > 0:
\quad
\arg\frac{m_{k}-m_{0}}{m_{1}-m_{0}} \in \left( 0,\frac{\pi}{2} \right)
\,,
\\
& \nu \le 0:
\quad
\arg\frac{m_{k}-m_{0}}{m_{1}-m_{0}} \in \left( \frac{\pi}{2},\pi \right)
\,,
\end{aligned}
\end{equation}
\begin{equation}
\label{second bound state constraint root}
\begin{aligned}
2:
\quad
& \nu \ge 0:
\quad
\arg\frac{m_{k}-m_{1}}{m_{1}-m_{0}} \in \left( 0,\frac{\pi}{2} \right)
\,,
\\
& \nu < 0:
\quad
\arg\frac{m_{k}-m_{1}}{m_{1}-m_{0}} \in \left( \frac{\pi}{2},\pi \right)
\,.
\end{aligned}
\end{equation}
Since the bound states with $\nu\ne 0$ do not exist at strong coupling, they appear at weak coupling if exactly one of the two walls (\ref{first wall root 2}, \ref{second wall root 2}) is crossed.
The states with $\nu=0$, which exist at strong coupling, appear at weak coupling if either none or both walls (\ref{first wall root 2}, \ref{second wall root 2}) are crossed.
For all three cases, $\nu>0$, $\nu<0$, and $\nu=0$, this means that the bound states in (\ref{first process root}, \ref{second process root}) exist semiclassically when the following condition is satisfied:
\begin{equation}
\label{bound state constraint root}
0 < \re\frac{m_{k}-m_{0}}{m_{1}-m_{0}} < 1
\end{equation}
(again, the inequality is strict because of (\ref{central charge dyon change root})).
This is precisely the semiclassical constraint derived in \cite{DHT} from first principles.
Analogously, if (\ref{bound state constraint root}) holds for $p$ different indices $k$ and $\tilde k$, there are towers of bound states with $p$ quarks (\ref{bound state general root}) having $j_{k}=1$ and $\tilde j_{\tilde k}=1$ for these indices and $j_{i}=0$ and $\tilde j_{\tilde i}=0$ for all other $i$ and $\tilde i$, in accordance with \cite{DHT}.

\paragraph{}

\begin{figure}[ht]
\centering
\includegraphics[width=65mm]{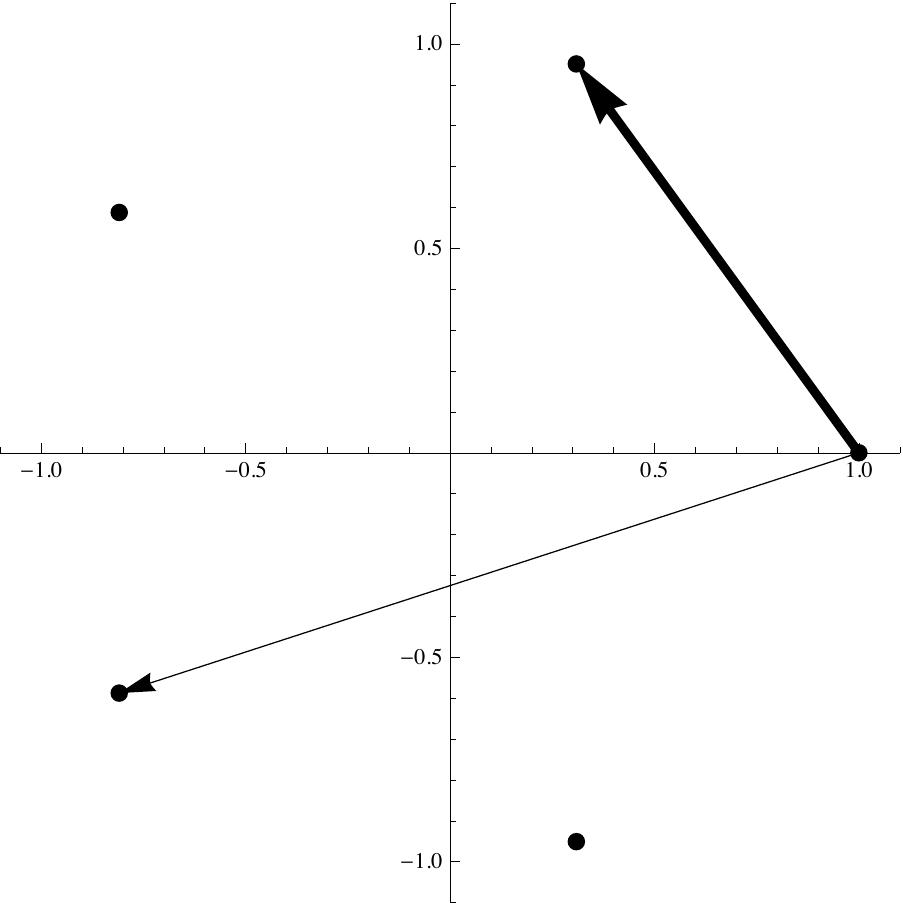}
\caption{
A dyon (thick vector connecting $m_{0}$ and $m_{1}$, representing its central charge near the massless point at strong coupling for $\nu>0$) and the quark that can form bound states with it (thin vector connecting $m_{0}$ and $m_{3}$, equal to its central charge) in the case of $\mathbb{Z}_{5}$-symmetric masses.
}
\label{fig: states root}
\end{figure}

\begin{figure}[H]
\centering
\includegraphics[height=65mm]{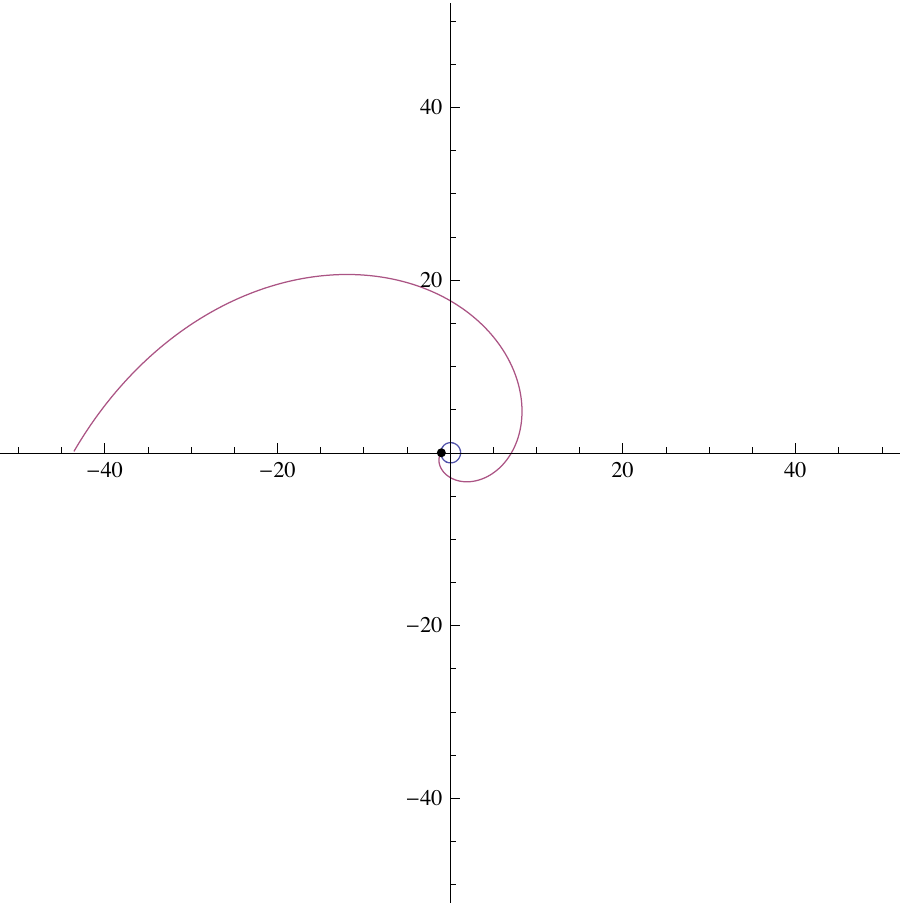}
\qquad
\includegraphics[height=65mm]{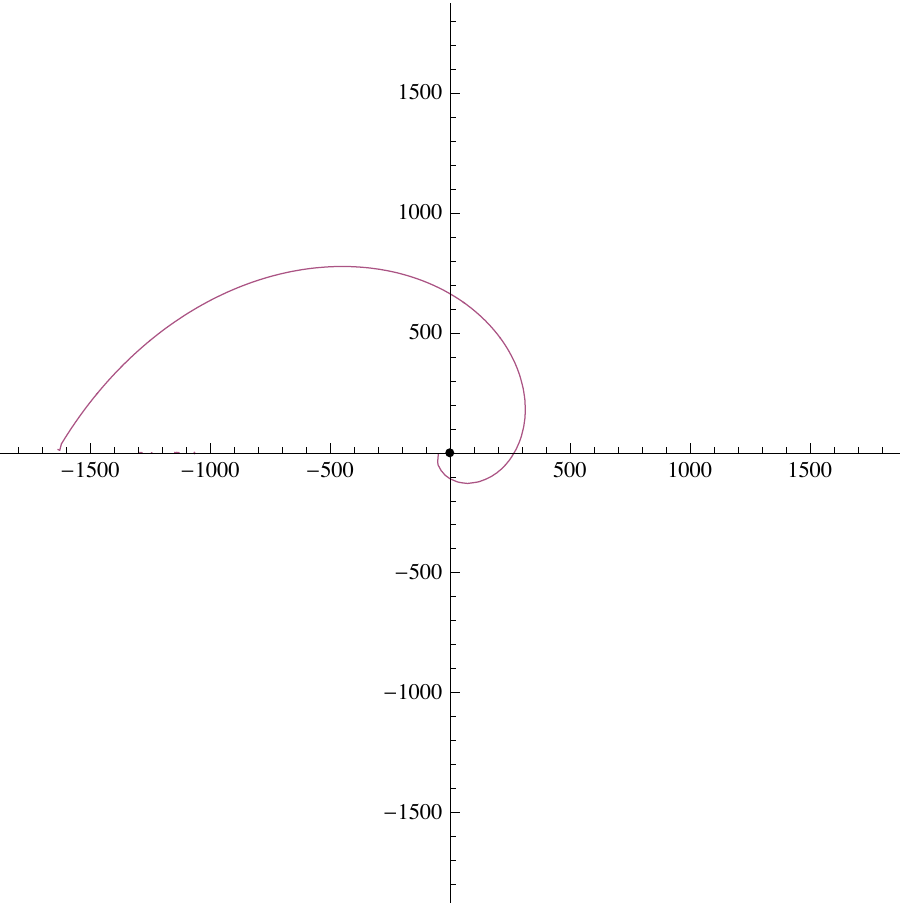}
\\[10pt]
\includegraphics[height=80mm]{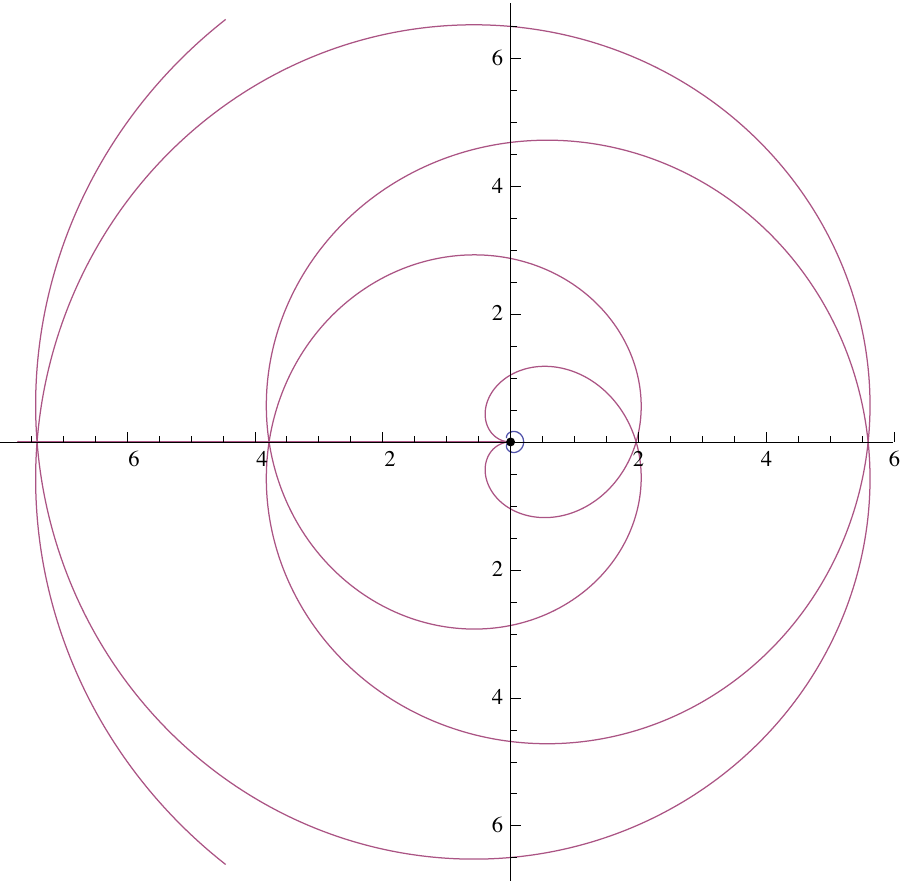}
\caption{
First row: secondary walls of marginal stability for $\mathbb{Z}_{3}$-symmetric masses in the $m_{0}^{3}/|m_{0}|^{2}$ plane for $\nu=1$ and $\nu=2$ (the walls for $\nu=-1$ and $\nu=-2$ are their reflections across the real axis); second row: the two spirals and the primary wall with all radii scaled as $r\to\log r$.
The plots correspond to the sector in figure \ref{fig: massless states root}.
}
\label{fig: walls secondary root}
\end{figure}

Applying this result to $\mathbb{Z}_{n}$-symmetric masses, it is easy to find the bound states in the weak-coupling limit.
The constraint (\ref{bound state constraint root}) reduces to
\begin{equation}
0 < \re\frac{e^{2\pi ki/n}-1}{e^{2\pi i/n}-1} = \re\frac{e^{2\pi(k-1/2)i/n}-e^{-\pi i/n}}{2i\sin(\pi/n)} < 1
\quad \iff \quad
-1 < \frac{\sin\frac{2\pi k-\pi}{n}}{\sin\frac{\pi}{n}} < 1
\,.
\end{equation}
Here, $(2\pi k-\pi)/n$ is a multiple of $\pi/n$, therefore, the inequality holds only for $(2\pi k-\pi)/n=\pi$, that is, for $k=(n+1)/2$.
For $\mathbb{Z}_{2l+1}$-symmetric masses with $l\in\mathbb{N}$, this means that only the bound states formed by one quark with $\gamma_{e\,(l+1)}=1$ are present (figure \ref{fig: states root}); for $\mathbb{Z}_{2l}$-symmetric masses, there are no bound states.
We can go back to equations (\ref{first bound state constraint root}, \ref{second bound state constraint root}) to find out which secondary walls of marginal stability exist in the case of $\mathbb{Z}_{2l+1}$-symmetric masses: (\ref{first process root}) is realised for $\nu>0$, (\ref{second process root}) is realised for $\nu<0$, and the corresponding walls are determined by (\ref{first wall root}) with $\nu>0$ and (\ref{second wall root}) with $\nu<0$ (plotted for $\mathbb{Z}_{3}$ in figure \ref{fig: walls secondary root}); in addition, all bound states with $\nu=0$ not belonging to the tower of bound states decay between strong- and weak-coupling, as discussed above.

\paragraph{}

The coils corresponding to $\nu=\pm(p+1)$ and to $\nu=\pm p$ for $p\ne 0$ are consecutive sections of the same spiral.
The clockwise and the counterclockwise spirals contain the coils with $\nu>0$ and $\nu<0$, respectively.
This follows from the fact that the rotation by $2\pi$ in the moduli space changes the electric charges by $1$.

\paragraph{}

The wall-crossing formula discussed above places strong constraints on the presence of any extra BPS states and their possible decay processes.
For example, in the four-dimensional model, additional states not belonging to the secondary tower with quantum numbers
\begin{equation}
((-\nu,\nu,0,0,\dots,0,1,0,0,\dots),(-1,1,0,0,0,\dots))
\,,
\end{equation}
analogous to the extra towers of state in the two-dimensional model discussed in \cite{BSY}, are not present.
In particular, the wall-crossing formula certainly forbids the obvious simultaneous decay process for the tower of such states into the known quarks and dyons of the model:
\begin{equation}
\begin{aligned}
((-\nu,\nu,1,0,0,\dots),(-1,1,0,0,0,\dots))
\overset{?}{\leftrightarrow} ((0,0,1,0,0,\dots),(-1,1,0,0,0,\dots))
\\
+\nu((-1,1,0,0,0,\dots),(0,0,0,0,0,\dots))
\,.
\end{aligned}
\end{equation}
To see this, suppose that the spectrum on right-hand side is correct, and we cross the wall in the other direction.
The symplectic product of the two charges on the right-hand side is two, which means that the formula (\ref{WCF pure}) should apply.
One can then check, however, that this leads to a different decay process into states having magnetic charges greater than one (in one $SU(2)$ subgroup), which are certainly absent from the model.

\paragraph{}

In this paper, we do not consider the corresponding two-dimensional theory directly~\footnote{
Note that a similar wall-crossing formula is believed to hold for 2d models of this type \cite{GMN 2d4d}.
}.
However, our spectrum in the 4d theory for $\mathbb{Z}_{n}$-symmetric masses coincides~\footnote{
More precisely, the two spectra coincide up to minor discrepancies which originate in the precise assignment of charges to the particles in the strong coupling region.
} with the relevant 2d spectrum obtained in the forthcoming paper \cite{BSY2}, and this agreement provides further support for the 2d/4d correspondence of \cite{Dorey 1998, DHT}.

\subsection*{Acknowledgements}

We would like to thank P.~A.~Bolokhov, M.~Shifman, and A.~Yung for discussions and for providing us with a preliminary version of \cite{BSY2}.
ND acknowledges the hospitality of the Yukawa Institute of Theoretical Physics, Kyoto University, and of the High Energy Theory group at Caltech, where part of this work was completed.
KP is supported by a research studentship from Trinity College, Cambridge.



\begin{thebibliography}{99}


\bibitem{Witten phases}
  E.~Witten,
  ``Phases of N=2 theories in two dimensions,''
  Nucl.\ Phys.\  B {\bf 403}, 159 (1993)
  [\href{http://arxiv.org/pdf/hep-th/9301042}{arXiv:hep-th/9301042}].

\bibitem{Dorey 1998}
  N.~Dorey,
  ``The BPS spectra of two-dimensional supersymmetric gauge theories with twisted mass terms,''
  JHEP {\bf 9811}, 005 (1998)
  [\href{http://arxiv.org/pdf/hep-th/9806056}{arXiv:hep-th/9806056}].

\bibitem{DHT}
  N.~Dorey, T.~J.~Hollowood and D.~Tong,
  ``The BPS spectra of gauge theories in two and four dimensions,''
  JHEP {\bf 9905}, 006 (1999)
  [\href{http://arxiv.org/pdf/hep-th/9902134}{arXiv:hep-th/9902134]}.

\bibitem{KS}
  M.~Kontsevich and Y.~Soibelman,
  ``Stability structures, motivic Donaldson--Thomas invariants and cluster transformations,''
  \href{http://arxiv.org/pdf/0811.2435}{arXiv:0811.2435 [math.AG]}.

\bibitem{GMN}
  D.~Gaiotto, G.~W.~Moore and A.~Neitzke,
  ``Four-dimensional wall-crossing via three-dimensional field theory,''
  Commun.\ Math.\ Phys.\  {\bf 299}, 163 (2010)
  [\href{http://arxiv.org/pdf/0807.4723}{arXiv:0807.4723 [hep-th]}].

\bibitem{Hanany Hori}
  A.~Hanany and K.~Hori,
  ``Branes and N=2 theories in two dimensions,''
  Nucl.\ Phys.\  B {\bf 513}, 119 (1998)
  [\href{http://arxiv.org/pdf/hep-th/9707192}{arXiv:hep-th/9707192}].

\bibitem{Hanany Tong}
  A.~Hanany and D.~Tong,
  ``Vortices, instantons and branes,''
  JHEP {\bf 0307}, 037 (2003)
  [\href{http://arxiv.org/pdf/hep-th/9707192}{arXiv:hep-th/0306150}].

\bibitem{Hanany Tong 2}
  A.~Hanany and D.~Tong,
  ``Vortex strings and four-dimensional gauge dynamics,''
  JHEP {\bf 0404}, 066 (2004)
  [\href{http://arxiv.org/pdf/hep-th/9707192}{arXiv:hep-th/0403158}].

\bibitem{SY}
  M.~Shifman and A.~Yung,
  ``Non-abelian string junctions as confined monopoles,''
  Phys.\ Rev.\  D {\bf 70}, 045004 (2004)
  [\href{http://arxiv.org/pdf/hep-th/0403149}{hep-th/0403149}].

\bibitem{BSY}
  P.~A.~Bolokhov, M.~Shifman and A.~Yung,
  ``BPS spectrum of supersymmetric CP(N$-$1) theory with Z$\rm _N$ twisted masses,''
  Phys.\ Rev.\  D {\bf 84}, 085004 (2011)
  [\href{http://arxiv.org/pdf/1104.5241}{arXiv:1104.5241 [hep-th]}].

\bibitem{BSY2}
  P.~A.~Bolokhov, M.~Shifman and A.~Yung,
  ``2D--4D correspondence: towers of kinks versus towers of monopoles in N=2 theories,''
  to appear.

\bibitem{CV}
  S.~Cecotti and C.~Vafa,
  ``On classification of N=2 supersymmetric theories,''
  Commun.\ Math.\ Phys.\  {\bf 158}, 569 (1993)
  [\href{http://arxiv.org/pdf/hep-th/9211097}{arXiv:hep-th/9211097}].

\bibitem{Veneziano Yankielowicz}
  G.~Veneziano and S.~Yankielowicz,
  ``An effective lagrangian for the pure N=1 supersymmetric Yang--Mills theory,''
  Phys.\ Lett.\  B {\bf 113}, 231 (1982).

\bibitem{Hori Vafa}
  K.~Hori and C.~Vafa,
  ``Mirror symmetry,''
  \href{http://arxiv.org/pdf/hep-th/0002222}{arXiv:hep-th/0002222}.

\bibitem{DDDS}
  A.~D'Adda, A.~C.~Davis, P.~Di~Vecchia and P.~Salomonson,
  ``An effective action for the supersymmetric CP(n$-$1) model,''
  Nucl.\ Phys.\  B {\bf 222}, 45 (1983).

\bibitem{Argyres Douglas}
  P.~C.~Argyres and M.~R.~Douglas,
  ``New phenomena in SU(3) supersymmetric gauge theory,''
  Nucl.\ Phys.\  B {\bf 448}, 93 (1995)
  [\href{http://arxiv.org/pdf/hep-th/9505062}{arXiv:hep-th/9505062}].

\bibitem{Olmez Shifman}
  S.~Olmez and M.~Shifman,
  ``Curves of marginal stability in two-dimensional CP(N$-$1) models with Z$\rm _N$-symmetric twisted masses,''
  J.\ Phys.\ A  {\bf 40}, 11151 (2007)
  [\href{http://arxiv.org/pdf/hep-th/0703149}{arXiv:hep-th/0703149}].

\bibitem{Shifman Vainshtein Zwicky}
  M.~Shifman, A.~Vainshtein and R.~Zwicky,
  ``Central charge anomalies in 2D sigma models with twisted mass,''
  J.\ Phys.\ A  {\bf 39}, 13005 (2006)
  [\href{http://arxiv.org/pdf/hep-th/0602004}{arXiv:hep-th/0602004}].

\bibitem{GMN 2d4d}
  D.~Gaiotto, G.~W.~Moore and A.~Neitzke,
  ``Wall-crossing in coupled 2d-4d systems,''
  \href{http://arxiv.org/pdf/1103.2598}{arXiv:1103.2598 [hep-th]}.

\end{thebibliography}
\end{document}